\documentclass[a4paper,preprintnumbers,floatfix,pra,twocolumn,showpacs,notitlepage,longbibliography, aps]{revtex4-2}
\usepackage{graphicx} % Required for inserting images
\usepackage{color}
\begin{document}
\title{Iso-$\mu/T$ holographic entropy and its attractor for a strongly coupled quantum fluid}
\author{W. Barreto}
\email{wbarreto@ivic.gob.ve}
\affiliation{Centro de Física, Instituto Venezolano de Investigaciones Científicas, Apartado Postal 20632, Caracas 1020-A, Venezuela}
\date{\today}
\begin{abstract}
Numerical evidence has shown that a period of vanishing entropy production during far-from-equilibrium stages induces subsequent violations of the dominant energy condition in strongly coupled plasmas. This behavior also appears to hold for a Bjorken-expanding, hot and dense strongly coupled quantum fluid. In this context, it has been established that the chemical potential-to-temperature ratio ($\mu/T$) in the medium increases with higher initial charge density, $\rho_0$, and/or lower initial energy density, $\varepsilon_0$.
Here, we present a numerical method to evolve Bjorken R-charged plasmas by varying $(\varepsilon_0,\rho_0)$ in order to generate a curve that preserves a constant $\mu/T$. This allows us to address the case in which all Bekenstein–Hawking entropy densities evolve towards configurations with the same $\mu/T$. This approach enables a more direct comparison with the $\mathcal{N}=4$ SYM plasma case ($\mu/T=0$) and, consequently, provides clearer evidence of the correlation between entropy production and violations of the dominant energy condition.
Moreover, this procedure offers a characterization (albeit \textcolor{black}{numerical and} partial) of the iso$-\mu/T$ entropy density hydrodynamic attractor for hot and dense strongly coupled quantum fluids.
\end{abstract}
\maketitle
\section{Introduction}
Dynamical systems out of equilibrium are well described, among other physical observables, by the production of entropy. As these systems evolve toward equilibrium, their evolution \textcolor{black}{reveals valuable insight into the onset of hydrodynamic behavior.} Particularly in the field of heavy-ion collisions, the Bjorken flow is quite relevant \cite{b83} and well studied  using both weakly and strongly coupled QCD-based descriptions, in different models of relativistic fluids \cite{cy10}--\cite{dklmnpry22}.

The holographic duality \cite{m98}--\cite{w98b} is an alternative theoretical framework to study strongly interacting models. This approach allows the full far-from-equilibrium numerical simulations of real-time dynamics of some strongly interacting QFTs; these QFTs may involve finite temperature and chemical potential. Holographically, one can study the hydrodynamic behavior at a late time when the system is subject to different initial conditions, which represent far-from-equilibrium settings. At early times, the system is highly dependent on these initial conditions, but they typically converge to the same hydrodynamic attractor, depending on a few variables, including temperature and chemical potential of the medium close to thermodynamic equilibrium. The system loses memory, erasing information with production of entropy.

Sometimes, once a problem has been solved, pursuing a large number of numerical experiments and using new post-processing methods allows us to explore and discover new features or get new insights of a physical system from well established models.

A series of holographic studies \cite{rb22}, \cite{dbr26}-- \cite{dbr25},  reported a far-from-equilibrium hot and dense strongly coupled quantum fluid, which can be analyzed evolving several physical observables, namely the energy density $\varepsilon$, the pressure anisotropy $\Delta p$, the scalar condensate $\langle O_\phi\rangle$, the charge density $\rho$ and the entropy density $s$. Two of these studies were for a Bjorken expanding flow, strongly coupled $\mathcal{N}=4$ Supersymmetric Yang-Mills plasma charged under Abelian $U(1)$ subgroup of the Global $SU(4)$ R-Symmetry.  
The first study \cite{rb22}, the 1 R-Charge Black Hole (1RCBH) model, contains a phase diagram with a critical point; the second one \cite{dbr26}, was the 2 R-Charge Black Hole (2RCBH) model. One motivation for undertaking such studies was the numerical evidence of the violation of the classical dominant energy condition \cite{rnbdd21}, \cite{rbn22} for a SYM ($\mu/T=0$) plasma undergoing a Bjorken flow, and the strong evidence of the correlation between the pressure anisotropy and the entropy density when the violation takes place far from equilibrium. We have cumulative evidence of this behavior for Bjorken flows under 1RCBH and 2RCBH \cite{rb22}, \cite{dbr26}, and even for systems without hydrodynamization \cite{rb24}, \cite{dbr25}. It is interesting that the correlation between the anisotropic pressure and the production of entropy exists both in homogeneous isotropization far from equilibrium and close to equilibrium. When close to equilibrium, the behavior is clearly related to the lowest quasi-normal mode \cite{jm16}, \cite{jm20}, \cite{ab25}, which seems to be a universal behavior.

Here we present a new numerical exploration of the 1RCBH model under a Bjorken flow \cite{rb22} using a new protocol. Such a procedure can be implemented for other models, as the 1RCBH under homogeneous isotropization, the 2RCBH under a Bjorken flow or under homogeneous isotropization. Here we use the numerical solver output data of \cite{rb22}, in order to study the 1RCBH model under a Bjorken flow more deeply and gain insights. For that reason, the involved equations for the 1RCBH-Bjorken model (and algorithms to solve them) are not presented here, as they are identical. However, \textcolor{black}{to make it minimally self-contained} we recall here only the dominant energy condition \cite{rnbdd21}, \cite{rbn22}
$$
\varepsilon \ge 0, \;\;\;\;\; -1 \le \Delta p/\varepsilon \le 2,  
$$
\textcolor{black}{and the initial profile for the subtracted metric anisotropy coefficient
\begin{widetext}
$$B_s(\tau_0,u) = \Omega_1 \cos(\gamma_1 u) + \Omega_2 \tan(\gamma_2 u) + \Omega_3 \sin(\gamma_3 u) + \sum_{i=0}^{5}\beta_i u^i  +\, \frac{\alpha}{u^4} \left[-\frac{2}{3} \ln\left(1+ \frac{u}{\tau _0}\right) + \frac{2 u^3}{9 \tau_0^3} - \frac{u^2}{3 \tau _0^2}+\frac{2 u}{3 \tau _0}\right],$$
\end{widetext}
where $u$ is the inverse of the radial coordinate $r$ for the asymptotically AdS$_5$ spacetime, whose boundary is at $r\rightarrow\infty\, (u=0)$, where $\tau_0$ will be the usual initial proper time of the Bjorken flow; 
and the set of parameters as given in Table I of \cite{rb22}:
\begin{table}[h]
\centering
\begin{tabular}{|c||c|c|c|c|c|c|c|c|c|c|c|c|c|}
\hline
$B_s\#$ & $\Omega_1$ & $\gamma_1$ & $\Omega_2$ & $\gamma_2$ & $\Omega_3$ & $\gamma_3$ & $\beta_0$ & $\beta_1$ & $\beta_2$ & $\beta_3$ & $\beta_4$ & $\beta_5$ & $\alpha$ \\
\hline
\hline
01 & 0 & 0 & 0 & 0 & 0 & 0 & 0.5 & -0.5 & 0.4 & 0.2 & -0.3 & 0.1 & 1 \\
\hline
02 & 0 & 0 & 0 & 0 & 0 & 0 & -0.2 & -0.5 & 0.3 & 0.1 & -0.2 & 0.4 & 1 \\
\hline
03 & 0 & 0 & 0 & 0 & 0 & 0 & 0.1 & -0.4 & 0.3 & 0 & -0.1 & 0 & 1 \\
\hline
04 & 0 & 0 & 1 & 1 & 0 & 0 & 0 & 0 & 0 & 0 & 0 & 0 & 1 \\
\hline
05 & 0 & 0 & 0 & 0 & 0 & 0 & -0.2 & -0.5 & 0 & 0 & 0 & 0 & 1 \\
\hline
06 & 0 & 0 & 0 & 0 & 0 & 0 & -0.2 & -0.4 & 0 & 0 & 0 & 0 & 1 \\
\hline
07 & 0 & 0 & 0 & 0 & 0 & 0 & -0.2 & -0.6 & 0 & 0 & 0 & 0 & 1 \\
\hline
08 & 0 & 0 & 0 & 0 & 0 & 0 & -0.3 & -0.5 & 0 & 0 & 0 & 0 & 1 \\
\hline
09 & 0 & 0 & 0 & 0 & 1 & 8 & 0 & 0 & 0 & 0 & 0 & 0 & 1 \\
\hline
10 & 1 & 8 & 0 & 0 & 0 & 0 & -0.2 & -0.5 & 0 & 0 & 0 & 0 & 1 \\
\hline
11 & 0.5 & 8 & 0 & 0 & 0 & 0 & -0.2 & -0.5 & 0 & 0 & 0 & 0 & 1 \\
\hline
\end{tabular}
%\caption{Set of parameters for the initial profile of the subtracted metric anisotropy \eqref{eq:Bs0} analyzed in this work.}
\label{tabICs}
\end{table}}

In the next section we explain the new post-processing construction, the iso$-\mu/T$, and how it can improve the analysis of results, particularly when the dominant energy condition is violated from below. In that way the iso$-\mu/T$ could help us have a better (\textcolor{black}{numerical and} partial) notion of the entropy density hydrodynamic attractor for hot and dense strongly coupled quantum fluids. To the best of our knowledge, this attractor has not been previously reported in the literature. 

\section{A case for an iso$-\mu/T$}
In practice, each simulation of a hot and dense plasma under a Bjorken flow \cite{rb22} requires two physical parameters, the initial density of charge $\rho_0$ and the initial energy density $\varepsilon_0$, these two parameters fix $\mu/T$. Except for the purely thermal SYM which is a particular case ($\mu/T=0$), each R-charge plasma simulation runs for a different chemical potential because we fix one physical parameter ($\varepsilon_0$ or $\rho_0$) at a time. It would be desirable, and more intuitive, to run different simulations for the same $\mu/T$, in particular when the energy conditions are close to being violated. {\it We assume that there exists, in the space of parameters $(\varepsilon_0,\rho_0)$, a curve in which $\mu/T$ is constant.} To explore that assumption, we propose a protocol for the 1RCBH model, but it can be implemented massively if necessary for any studied Bjorken R-charged plasma. The procedure can also be useful in other contexts, such as the homogeneous isotropization \cite{rb24}, \cite{dbr25}. 

We run a large quantity of simulations in the space of parameters $(\varepsilon_0,\rho_0)$, selecting in post-processing the $\mu/T$ value of interest, interpolating and extrapolating, up to some specified numerical precision. This could require the implementation of automation to ameliorate cumbersome and heavy work. A simple approach to this automation is to bracket the edges of the region of parameters of interest, and then search within this space for the closest points that result in the desired value for $\mu/T$. If the value is not found, additional iterations of the process are run. This helps us build with some ease a series of curves with a given iso-$\mu/T$, while avoiding a larger search in the space, given that every iso-$\mu/T$ value calculation involves solving a 1RCBH model. 

This exploration is done for every initial condition that we report in \cite{rb22}, for the 1RCBH model, observing that $x=\mu/T$ and $x_c=\pi/\sqrt{2}$ is the critical point value.

\begin{figure}
\includegraphics[width=0.42\textwidth]{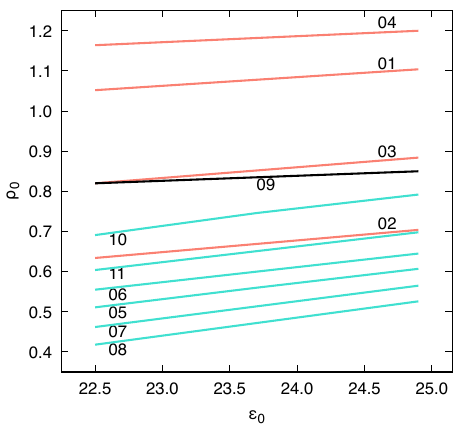}
\caption{Initial density of charge $\rho_0$ as a function of initial energy density $\varepsilon_0$; the isolines correspond to $x/x_c=0.442$ for all the initial conditions $B_s$ (Table I) and $\phi_s$01 (Table II) in Ref. \cite{rb22}. Observe that $x=\mu/T$ and $x_c=\pi/\sqrt{2}$ is the critical point value. The salmon curves correspond to initial conditions which do not violate the dominant energy condition; the turquoise curves correspond to initial conditions which violate the dominant eneegy condition; in black an anomalous case.}
\label{figure1}
\end{figure}

\begin{figure}
\includegraphics[width=0.42\textwidth]{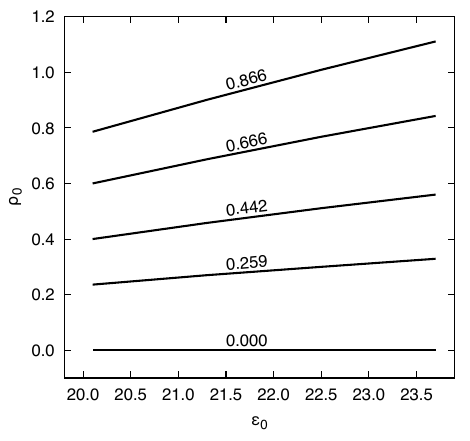}
\caption{Initial density of charge $\rho_0$ as a function of initial energy density $\varepsilon_0$; the isolines correspond to five values of $x/x_c$, for the initial conditions $B_s$05 and $\phi_s$01 (see Table I and Table II in Ref. \cite{rb22}). These isolines are qualitatively representative for any of the initial conditions. Observe that $x=\mu/T$ and $x_c=\pi/\sqrt{2}$ is the critical point value.}
\label{figure2}
\end{figure}
\begin{figure}
\center
{\includegraphics[width=0.4\textwidth]{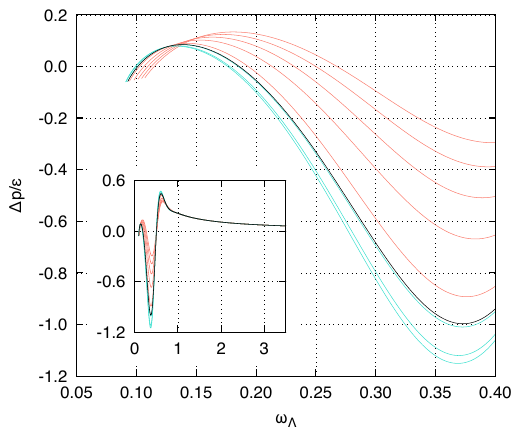}}
\caption{Early time windows for the normalized pressure anisotropy -- results obtained for variations $(\varepsilon_0,\rho_0)$ which fix $x/x_c=0.259$, for the initial conditions $B_s$05 and $\phi_s$01 (see Table I and Table II in Ref. \cite{rb22}). The black line corresponds to $(\varepsilon_0,\rho_0)=(20.46,0.246)$ and defines the transition between the salmon and turquoise curves, that is, the violation or not of the dominant energy condition. Observe that $x_c=(\mu/T)_c=\pi/\sqrt{2}$ is the critical point. The inset corresponds to the same observable for most of the evolution.}
\label{figure3}
\end{figure}
\begin{figure}
\center
{\includegraphics[width=0.4\textwidth]{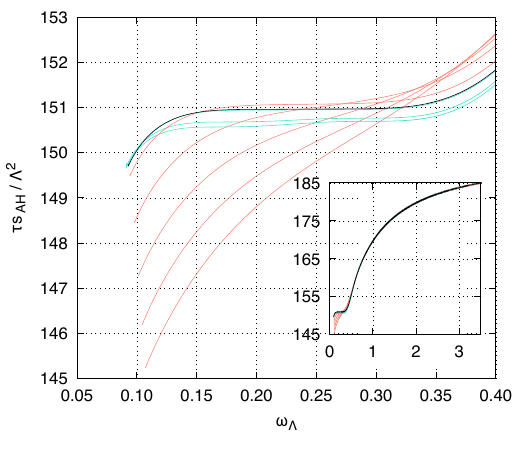}}
\caption{Early time windows for the non-equilibrium entropy density -- results obtained for variations $(\varepsilon_0,\rho_0)$ which fix $x/x_c=0.259$, for the initial conditions $B_s$05 and $\phi_s$01 (see Table I and Table II in Ref. \cite{rb22}). The black line corresponds to $(\varepsilon_0,\rho_0)=(20.46,0.246)$ and defines the transition between the salmon and turquoise curves, that is, the violation or not of the dominant energy condition. Observe that $x_c=(\mu/T)_c=\pi/\sqrt{2}$ is the critical point. The inset corresponds to the same observable for most of the evolution.}
\label{figure4}
\end{figure}
\begin{figure}
{\includegraphics[width=0.4\textwidth]{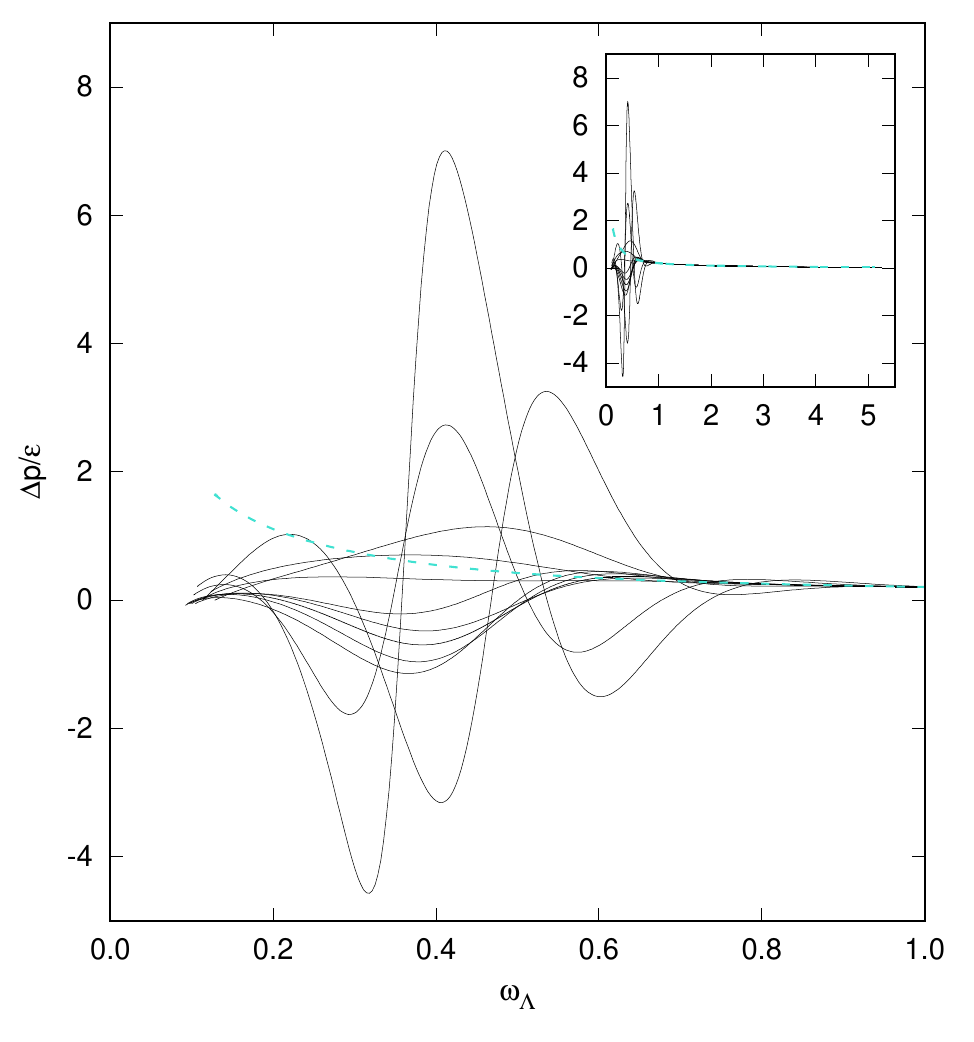}}
\caption{Evolution of the normalized pressure anisotropy for $\varepsilon_0=22.5$ and $\mu/T=0.442\,x_c$ and for all the initial conditions $B_s$ (Table I) and $\phi_s$01 (Table II) in Ref. \cite{rb22}. The dashed and turquoise line is the Navier-Stokes attractor. The inset corresponds to the same observable for the whole evolution.}
\label{figure5}
\end{figure}
\begin{figure}
{\includegraphics[width=0.4\textwidth]{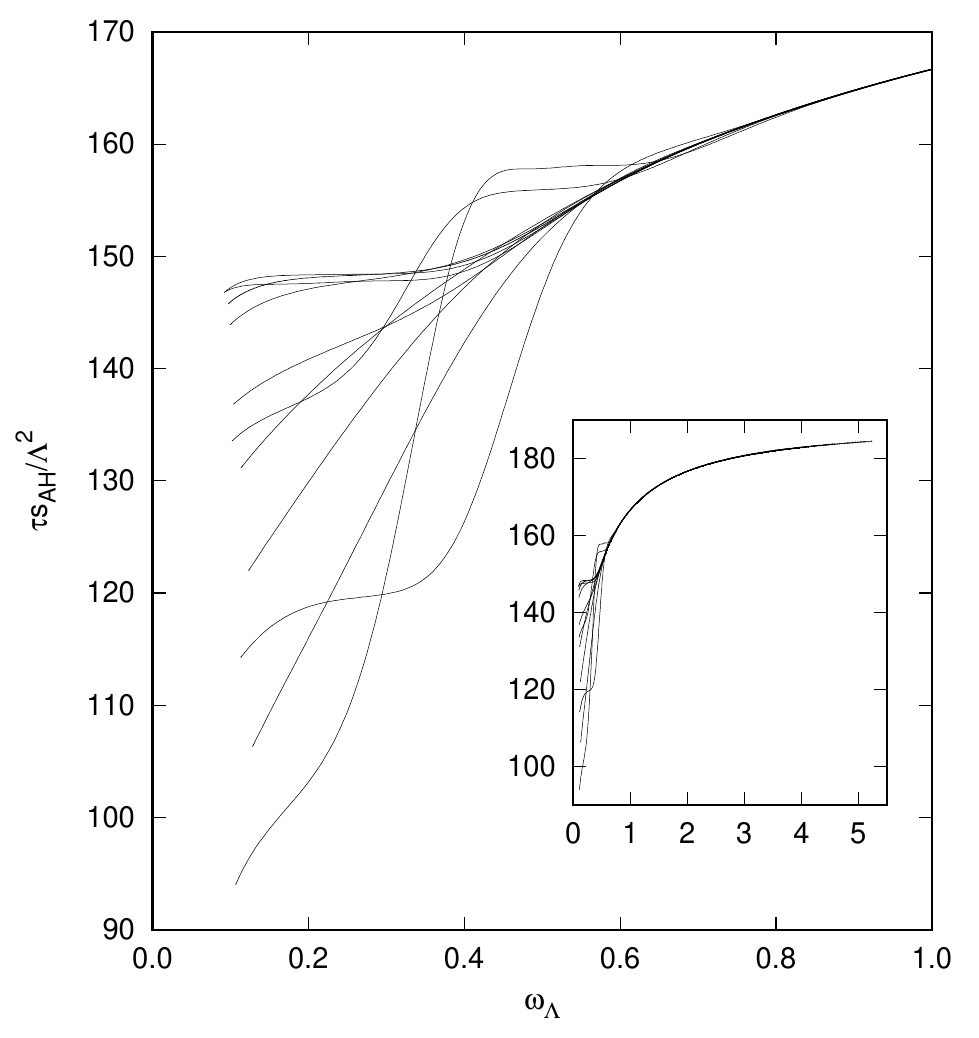}}
\caption{Evolution of the non-equilibrium entropy density for $\varepsilon_0=22.5$ and $\mu/T=0.442\,x_c$ and for all the initial conditions $B_s$ (Table I) and $\phi_s$01 (Table II) in Ref. \cite{rb22}. The inset corresponds to the same observable for the whole evolution. Observe that the unique curve should be very close to the hydrodynamic attractor.}
\label{figure6}
\end{figure}
\begin{figure}
{\includegraphics[width=0.4\textwidth]{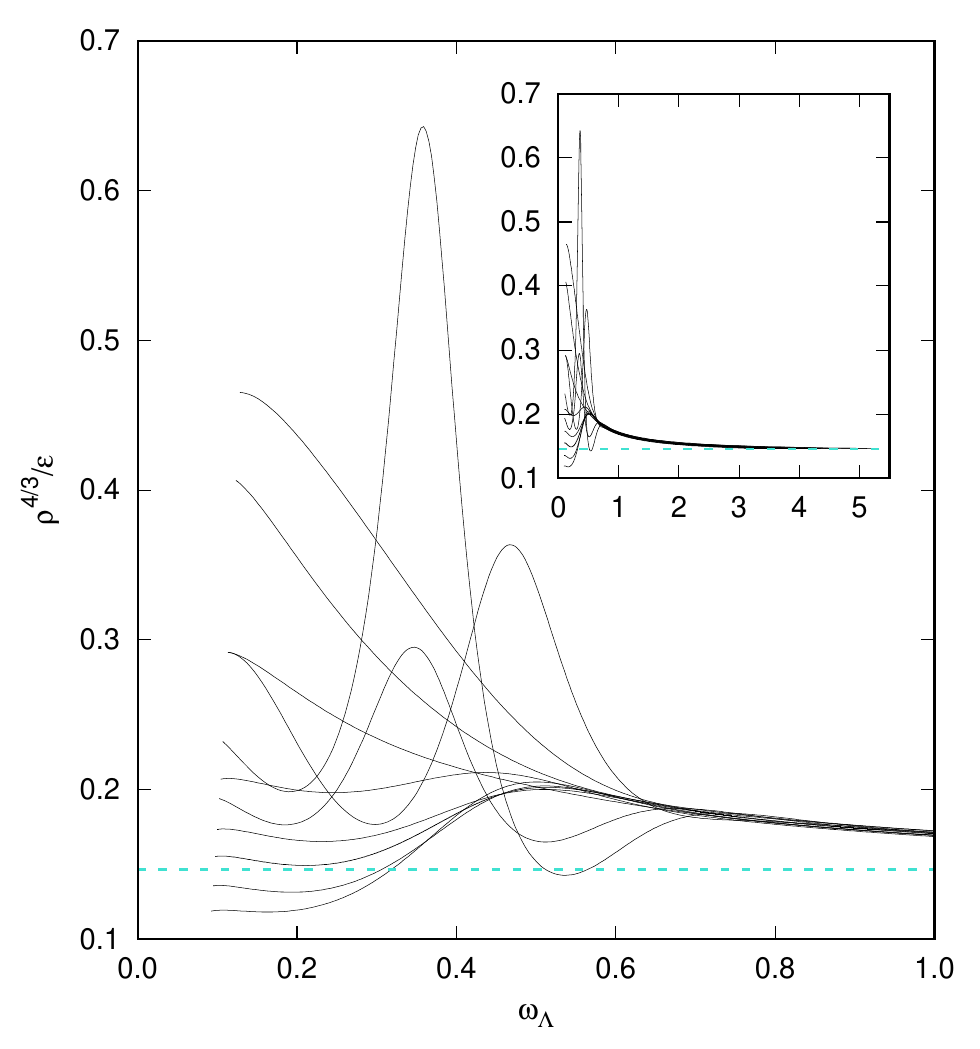}}
\caption{Evolution of the normalized charge density for $\varepsilon_0=22.5$ and $\mu/T=0.442\,x_c$ and for all the initial conditions $B_s$ (Table I) and $\phi_s$01 (Table II) in Ref. \cite{rb22}. The dashed turquoise line corresponds to the thermal equilibrium and the inset to the whole evolution of the same observable.}
\label{figure7}
\end{figure}
\begin{figure}
{\includegraphics[width=0.4\textwidth]{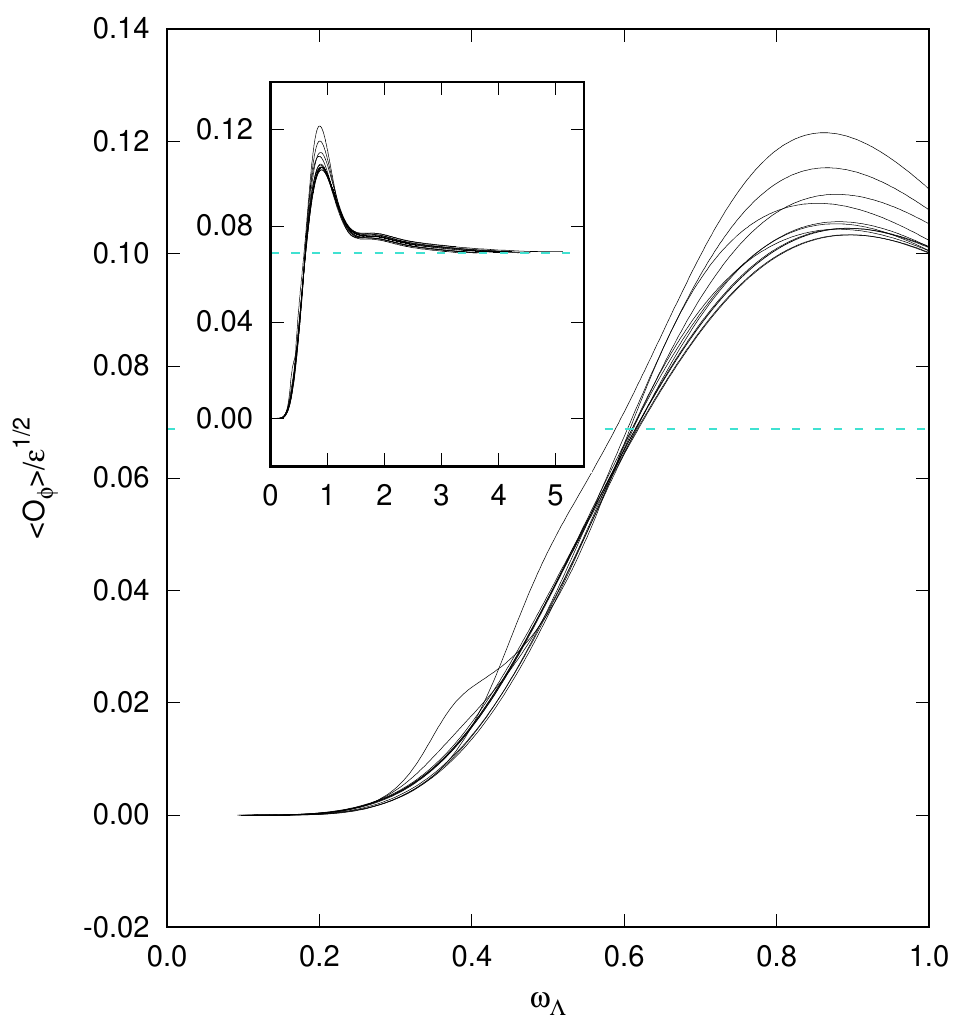}}
\caption{Evolution of the normalized scalar condensate for $\varepsilon_0=22.5$ and $\mu/T=0.442\,x_c$ and for all the initial conditions $B_s$ (Table I) and $\phi_s$01 (Table II) in Ref. \cite{rb22}. The dashed turquoise line corresponds to the thermal equilibrium and the inset to the whole evolution of the same observable.}
\label{figure8}
\end{figure}
\begin{figure}
{\includegraphics[width=0.4\textwidth]{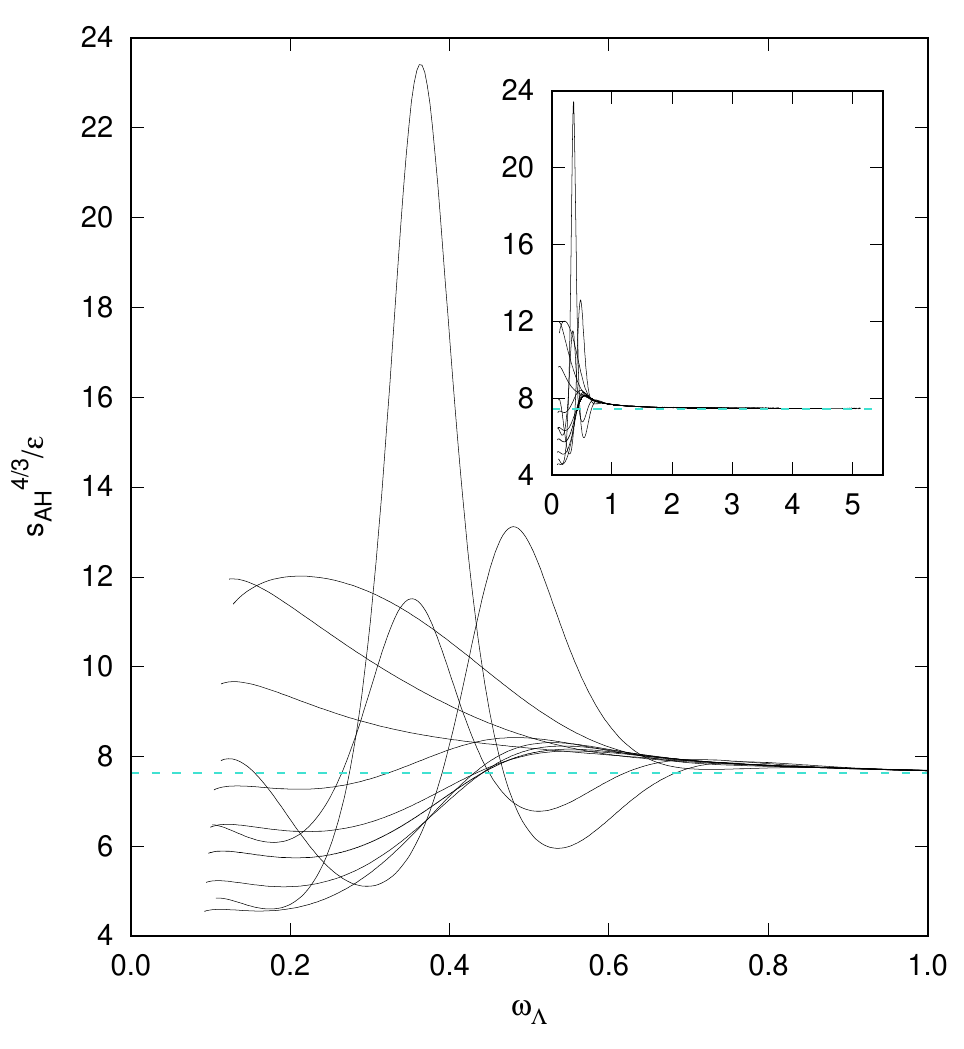}}
\caption{Evolution of the normalized non-equilibrium entropy density for $\varepsilon_0=22.5$ and $\mu/T=0.442\,x_c$ and for all the initial conditions $B_s$ (Table I) and $\phi_s$01 (Table II) in Ref. \cite{rb22}. The dashed turquoise line corresponds to the thermal equilibrium and the inset to the whole evolution of the same observable.}
\label{figure9}
\end{figure}
\begin{figure}
{\includegraphics[width=0.4\textwidth]{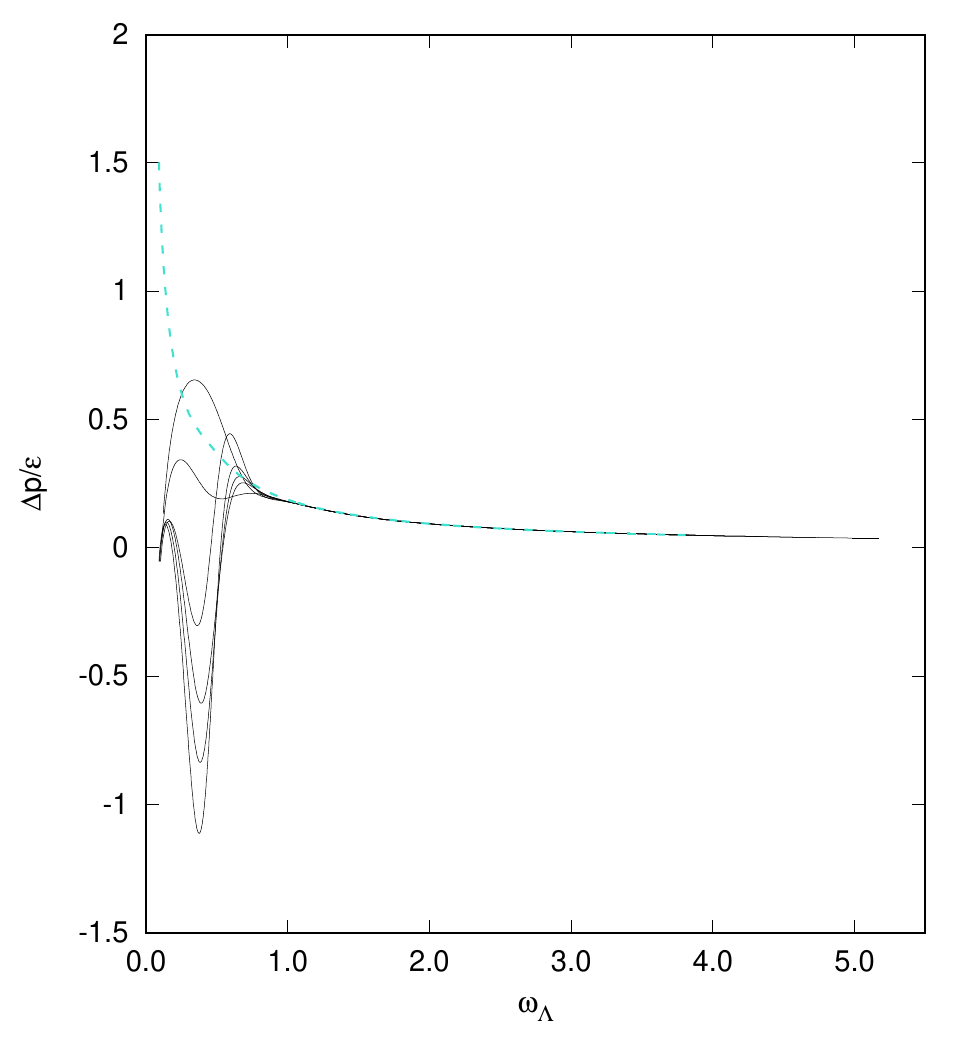}}
\caption{Evolution of the normalized pressure anisotropy for $\varepsilon_0=22.5$ and $\mu/T=0.848\,x_c$ and for six of the initial conditions $B_s$ (Table I) and $\phi_s$01 (Table II) in Ref. \cite{rb22}. The dashed and turquoise line is the Navier-Stokes attractor.}
\label{figure10}
\end{figure}
\begin{figure}
{\includegraphics[width=0.4\textwidth]{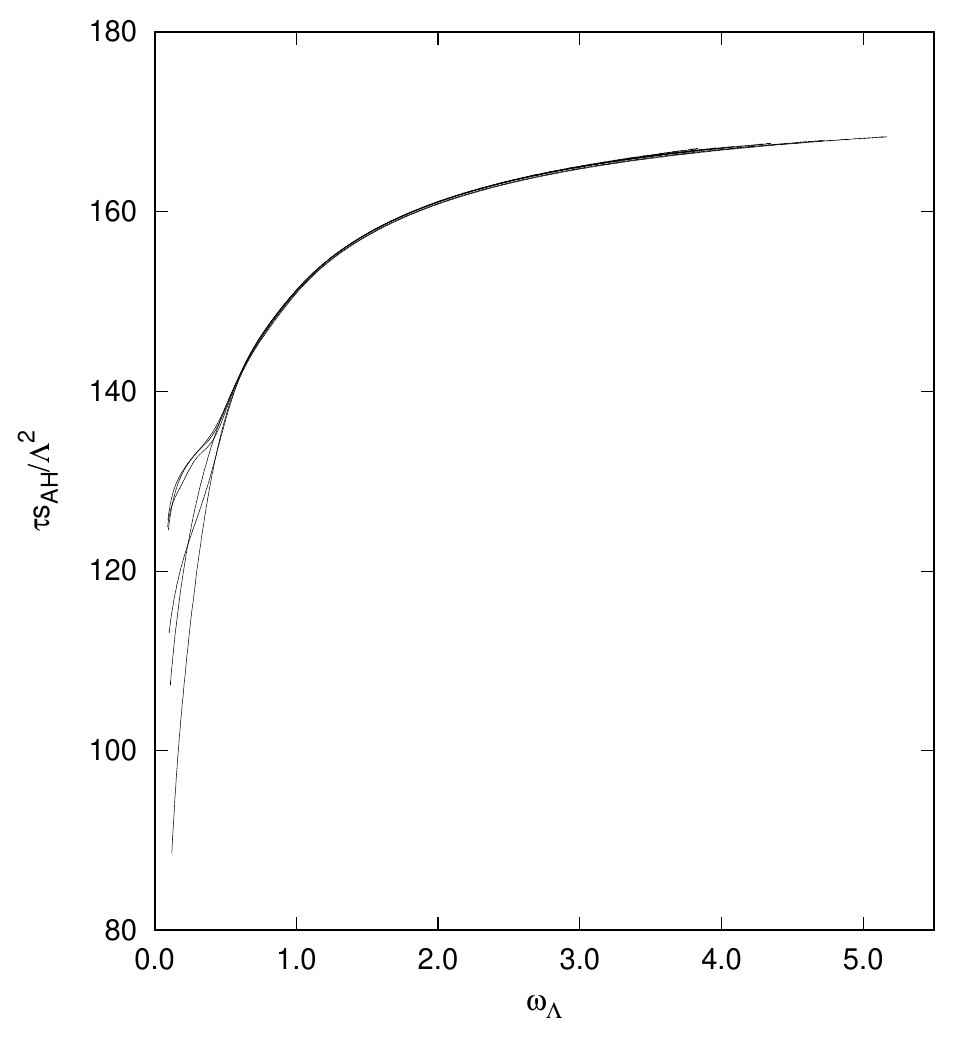}}
\caption{Evolution of the non-equilibrium entropy density for $\varepsilon_0=22.5$ and $\mu/T=0.848\,x_c$ and for six of the initial conditions $B_s$ (Table I) and $\phi_s$01 (Table II) in Ref. \cite{rb22}.}
\label{figure11}
\end{figure}

\section{Discussion of new results}
Following the iso$-\mu/T$ procedure for the production of new results, we consider for any run the initial profile of the subtracted dilaton field as $\phi_s(0,u)=0$. 
Figure \ref{figure1} displays the iso$-\mu/T$ lines for $x/x_c=0.442$ and   
for all the initial profiles of the subtracted metric anisotropy given in Table I of \cite{rb22}\textcolor{black}{, as recalled in section I}.
Figure \ref{figure2} shows five iso$-\mu/T$ lines for the initial profile of the subtracted metric anisotropy $B_s$05. Figures \ref{figure3} and \ref{figure4} show the normalized anisotropy pressure and the non-equilibrium entropy density, respectively,  close to the dominant energy condition violation, for the initial profile of the subtracted metric anisotropy $B_s$05, and $\mu/T=0.259\,x_c$.
Figures \ref{figure5}, \ref{figure6}, \ref{figure7}, \ref{figure8} and \ref{figure9} show the evolution of all the observables, for all the initial profiles of the subtracted metric anisotropy, for $\mu/T=0.442\,x_c$, and $\varepsilon_0=22.5$. Each initial profile requires a different value of $\rho_0$ to keep the iso-$\mu/T$ (first point of each isoline in figure \ref{figure1}). Figures \ref{figure10} and \ref{figure11} show the normalized pressure anisotropy and the non-equilibrium entropy density for some initial profiles of the subtracted metric anisotropy, for $\mu/T=0.848\,x_c$, and $\varepsilon_0=22.5$.

The first observation is that the iso$-\mu/T$ are initial data dependent, as is clearly shown in Figure \ref{figure1}. Interestingly, we observe that that the initial condition profile which violates the dominant energy condition from below has a higher $d\rho_0/d\varepsilon_0$, with the exception of the initial condition $B_s$09. 
\textcolor{black}{This apparent anomalous initial condition exhibits the peculiar behavior that its normalized pressure anisotropy oscillates differently compared with the other cases that violate the dominant energy condition before hydrodynamization.}
\textcolor{black}{In its last oscillation (below the attractor), the dominant energy condition is not violated from below; correspondingly, the entropy density does not form a plateau. This anomalous behavior appears to be reflected in an even lower slope $d\rho_0/d\varepsilon_0$ along its iso-$\mu/T$
curve than that found for the initial conditions which do not violate the dominant energy condition from below.}

\textcolor{black}{For the same initial condition the slope $d\rho_0/d\varepsilon_0$ increases for $\mu/T$ from $0$ to $\pi/\sqrt{2}$ (see Figure \ref{figure2}); this observation holds for all initial conditions considered. }

%For the same initial condition the slope $d\rho_0/d\varepsilon_0$ increases for $\mu/T$ from $0$ to $\pi/\sqrt{2}$ (see Figure \ref{figure2}); this observation holds for any considered initial condition. 

One can think, considering the first observation and previous results \cite{rnbdd21}, \cite{rbn22}, that for a given initial condition and a given iso-$\mu/T$, exists a critical pair $(\varepsilon_0,\rho_0)$ for which the dominant energy condition is violated. This is clearly observed in figures \ref{figure3} and \ref{figure4} for the initial condition $B_s$05; for the iso-$\mu/T=0.259\,x_c$ the critical pair is $(20.46,0.246)$. Just for comparison and for the same initial condition, for the iso-$\mu/T=0$ the critical pair is $(20.49,0.000)$ (see panels  (a) and (b) of figure 28 in \cite{rb22}). Now, with the iso-$\mu/T(=0.259\,x_c)$ protocol we have a stronger evidence of the correlation between the pressure anisotropy and the entropy density for $\mu/T\ne 0$ as displayed in figures \ref{figure3} and \ref{figure4}. The iso-$\mu/T$ procedure allows us to compare with the $\mathcal{N}=4$ SYM plasma case ($\mu/T=0$) \cite{rnbdd21}, where evidence of violation of the dominant energy condition was first reported. The entropy density forms a plateau when the pressure anisotropy is at the threshold of the dominant energy condition violation from below ($\Delta p/\varepsilon\approx -1)$.   

The most interesting results are displayed in Figures \ref{figure5}-\ref{figure9}, that is, the observables of the studied system. As in the case $\mu/T=0$ (pure thermal SYM) \cite{rnbdd21}, \cite{rbn22}, all the considered initial data with the same $\mu/T=0.442$ evolve toward the same equilibria for late times. For all the considered initial conditions the normalized pressure anisotropy goes to the Navier-Stokes regime, as expected. The entropy density goes to its hydrodynamic attractor for a hot and dense strongly interacting plasma, which marks the hydrodynamization. It is remarkable that for the selected iso-$\mu/T$ we do not observe evidence of delay in the hydrodynamization compared with the $\mathcal{N}=4$ SYM plasma case ($\mu/T=0$), except that the entropy density values are diminished in the same interval of time. The entropy density hydrodynamizes earlier than the normalized pressure anisotropy, but this could be due to the  normalization of the observables. We discuss this point later. \textcolor{black}{The most striking observation} at this point about the entropy density is the clear evidence of an earlier unique curve which should be very close to the hydrodynamic attractor for the iso-$\mu/T$ considered. 
% This will happen for any iso-$\mu/T$ and any initial condition.
This hydrodynamical attractor is not known in the literature for the hot and dense strongly interacting Bjorken flow. The normalized density of charge for all the initial data is observed to thermalize  but not forming a unique curve as the pressure anisotropy or the entropy density. 
%, as expected, because the charge conservation is fixed at all times by the U(1) symmetry and not by gradient-resummed hydrodynamics. 

The normalized scalar condensate behaves in the same way, with an even late equilibration
%, by a different reason. There is not a conserved current or symmetry for the scalar condensate. It evolves according to its dynamical equation of motion and relaxes 
relaxing toward its thermal equilibrium value 
%primarily through a quasinormal-mode decay, 
not forming a unique curve as well. The normalized entropy density clearly does not go to the same curve as early as the entropy density shown in figure \ref{figure6}. It behaves like the normalized pressure anisotropy. For that reason we conjecture that the delay in entering the unique curve is due to the normalization. At least for the entropy density we could confirm here that it hydrodynamizes earlier when we increase the chemical potential. But we found that increasing the chemical potential (see figure \ref{figure11}) and following the iso-$\mu/T$ protocol, the entropy density for the hot and dense plasma does not form an unique curve, in this sense delaying the hydrodynamization as has been reported in \cite{rb22}, \cite{dbr26}. \textcolor{black}{However, it is noteworthy that the different initial data result in memory loss and early (about $\omega_\Lambda \approx 0.6$) equilibration.} One may think it is a consequence of the precision work ($10^{-3}$) for the iso-$\mu/T$ runs procedure, but we have numerical evidence that it is not the case. Finally, two additional observations are in order. First, the pressure anisotropy hydrodynamizes early when we increase the chemical potential (see Figure \ref{figure10}). Second, the charge density is strongly correlated with the entropy density when displayed normalized (see figures \ref{figure7} and \ref{figure9}). This last observation is noticeable because the charge density for the 1RCBH model is determined by $\rho=\rho_0/\tau$ and one can think that nothing interesting could happen. However, also the presence of an attractor for the charge density can be inferred. From our results here\textcolor{black}{,} we cannot say the same for the scalar condensate. 

%\newpage
\section{Conclusions}

\textcolor{black}{In this work we studied the 1RCBH model, as reported in \cite{rb22}, with a new perspective. We explore how we can gain physical insight just running a large quantity of initial data exploring the space of parameters with the target iso-$\mu/T$. No other method is required in post-processing except an efficient inter(extra)-polation up to the desired precision; in the present study of order $10^{-3}$.}

\textcolor{black}{Following this protocol, we find that the normalized entropy density from all considered initial conditions collapses onto a single curve at low to moderate $\mu/T$, providing the first (partial) \textcolor{black}{numerical} characterization of the entropy density hydrodynamic attractor for a hot and dense strongly coupled quantum fluid, to the best of our knowledge not previously reported in the literature. This near-universal curve breaks down at higher $\mu/T$, where hydrodynamization is delayed and the curves fail to converge onto a single curve. Nonetheless, the initial data appear to lose memory and approach equilibration earlier, even though full hydrodynamization is delayed. Whether this loss of memory without full collapse reflects a genuine feature of the attractor at large $\mu/T$, or a limitation of the $10^{-3}$ precision employed here, remains an open question which \textcolor{black}{requires analytical results or} demands extra computational resources. We explored the last possibility up to $10^{-4}$ without any change in our conclusion. Regardless of the required precision, each run for a specific point in the space of parameters takes 3 minutes on a MacBook M1 Pro and 16 GB of memory.  We would like to stress that the Borel attractor is known only for the pressure anisotropy with $\mu/T=0$ and the needed reported precision is of $10^{-100}$ order, using $450$ digits and 240 terms in the gradient expansion \cite{hjw13}.}

\textcolor{black}{When the dominant energy condition is going to be violated -before hydrodynamization- the entropy production transiently goes to zero, even for a hot and dense strongly interacting plasma.}

\textcolor{black}{We also observe that the normalized charge density closely tracks the normalized entropy density during the approach to equilibrium, despite the trivial power-law decay $\rho=\rho_0/\tau$ of the unnormalized quantity; this suggests that an attractor-like behavior may also be present for the \textcolor{black}{energy} density. In addition, we find that the entropy density hydrodynamizes earlier than the pressure anisotropy under the iso-$\mu/T$ protocol.}

\textcolor{black}{Finally, the iso-$\mu/T$ protocol introduced here can be extended to other holographic settings, such as the 2RCBH model under Bjorken flow \cite{dbr26}, or to homogeneous isotropization scenarios for both the 1RCBH and 2RCBH models \cite{rb24}, \cite{dbr25}. Beyond the formal holographic setting, understanding how the hydrodynamic attractor is modified at finite chemical potential is conceptually relevant to the beam-energy-scan program at RHIC and upcoming experiments at FAIR and NICA \cite{sy20}, \cite{almaaloletal22}, \cite{pw23}, \cite{lns26}.}

\acknowledgments
The author acknowledges financial support by the National Council for Scientific and Technological Development (CNPq) under grant number 407162/2023-2. He is grateful for the warm hospitality of the Low Temperature Laboratory Group at IVIC during the development of this work. The author also thanks to \textcolor{black}{Rômulo Rougemont for insightful questions and comments regarding some of our results}; Nairy Villarreal, Luis Araque and {\textcolor{black}{Luis Rosales} for comments on the original version. The iso-$\mu/T$ bracketing and search  algorithm was partially cross-checked using a code written with the assistance of Claude Sonnet 5 AI.


\thebibliography{99}
\bibitem{b83} J. Bjorken, Phys. Rev. D 27, 140 (1983).
\bibitem{cy10} P. Chesler, L. Yaffe, Phys. Rev. D 82, 026006 (2010).
\bibitem{hjw12a} M. Heller, R. Janik, P. Witaszczyk, Phys. Rev. Lett. 108, 201602 (2012).
\bibitem{hjw12b} M. Heller, R. Janik, P. Witaszczyk, Phys. Rev. D 85, 126002 (2012).
\bibitem{jps14} J. Jankowski, G. Plewa, M. Spalinski, JHEP 12, 105 (2014).
\bibitem{p14} J. Pedraza, Phys. Rev. D 90(4), 046010 (2014).
\bibitem{bcdg15} L. Bellantuono, P. Colangelo, F. De Fazio, F. Giannuzzi, JHEP 07, 053 (2015).
\bibitem{hkss18} M. Heller, A. Kurkela, M. Spalinski, V. Svensson, Phys. Rev. D 97(9), 091503 (2018).
\bibitem{r18} P. Romatschke, Phys. Rev. Lett. 120 (1), 012301 (2018). 
\bibitem{dgpy17} B. DiNunno, S. Grozdanov, J. Pedraza, S. Young, JHEP 10, 110 (2017).
\bibitem{s18} M. Spalinski, Phys. Lett. B 776, 468 (2018).
\bibitem{snd18} M. Strickland, J. Noronha, G. Denicol, Phys. Rev. D 97(3), 036020 (2018).
\bibitem{cgm18} J. Casalderrey-Solana, N.  Gushterov, B. Meiring, JHEP 04, 042 (2018). 
\bibitem{crn19} R. Critelli, R. Rougemont, J. Noronha, Phys. Rev. D 99 (6), 066004 (2019).
\bibitem{kvww20} A. Kurkela, W. van der Schee, U. Wiedemann, B. Wu, Phys. Rev. Lett. 124 (10), 102301 (2020).
\bibitem{aks20} D. Almaalol, A. Kurkela, M. Strickland, Phys. Rev. Lett. 125(12), 122302 (2020). 
\bibitem{dnm20} T. Dore, J. Noronha-Hostler, E. McLaughlin, Phys. Rev. D 102(7), 074017 (2020).
\bibitem{rnbdd21} R. Rougemont, J. Noronha, W. Barreto, G. Denicol, T. Dore, Phys. Rev. D 104 (12), 126012 (2021).
\bibitem{rbn22} R. Rougemont, W. Barreto, J. Noronha, Phys. Rev. D 105 (4), 046009 (2022).
\bibitem{cks22} C. Cartwright, M. Kaminski, B. Schenke, Phys. Rev. C 105 (3), 034903 (2022).
\bibitem{rb22} R. Rougemont, W. Barreto, Phys. Rev. D 106 (12), 126023 (2022).
\bibitem{ckk23} C. Cartwright, M. Kaminski, M. Knipfer, Phys. Rev. D 107 (10), 106016 (2023).
\bibitem{dklmnpry22} T. Dore, J. Karthein, I. Long, D. Mroczek, J. Noronha-Hostler, P. Parotto, C. Ratti, Y. Yamauchi, Phys. Rev. D 106 (9), 094024 (2022).
\bibitem{m98} J. Maldacena, Adv. Theor. Math. Phys. 2, 231 (1998). 
\bibitem{gkp98} S. Gubser, I. Klebanov, A. Polyakov, Phys. Lett. B 428, 105 (1998).
\bibitem{w98a} E. Witten, Adv. Theor. Math. Phys. 2, 253 (1998).
\bibitem{w98b} E. Witten, Adv. Theor. Math. Phys. 2, 505 (1998).
\bibitem{dbr26} G. de Oliveira, W. Barreto, R. Rougemont, to appear in EPJC (2026).
\bibitem{rb24} R. Rougemont and W. Barreto, Phys. Rev. D 109, 126009 (2024).
\bibitem{dbr25} G. de Oliveira, W. Barreto, and R. Rougemont, J. High Energ. Phys. 10 (2025) 106.
\bibitem{jm16} A. Jansen, J. Magan, Phys. Rev. D 94, 104007 (2016).
\bibitem{jm20} A. Jansen, B. Meiring, Phys. Rev. D 101, 126012 (2020).
\bibitem{ab25} L. Araque, W. Barreto, Phys. Rev. D 111, 106014 (2025).
\bibitem{hjw13} M. Heller, R. Janik, P. Witaszczyk, Phys. Rev. Lett. 110, 211602 (2013).
\bibitem{sy20} C. Shen, L. Yan, Nucl. Sci. Tech. 31, 122 (2020).
\bibitem{almaaloletal22} D. Almaalol et al., arXiv:2209.05009 [nucl-ex]
\bibitem{pw23} L. Pang, X. Wang, Nucl. Sci. Tech. 34, 194 (2023).
\bibitem{lns26} J. Lakshmi, V. Naik, J. Sreekanth, Phys. G, 53, 075106 (2026)
\end{document}